\documentclass{emulateapj}

\shorttitle{Planet formation in binary systems}

\begin{document}

\title{Planet formation in binary systems: A separation-dependent
  mechanism?}

\author{G. Duch\^ene\altaffilmark{1}} \affil{Astronomy Department,
  University of California, Berkeley, CA 94720-3411, USA}
\email{gduchene@berkeley.edu}

\altaffiltext{1}{Universit\'e Joseph Fourier - Grenoble 1 / CNRS,
  Laboratoire d'Astrophysique de Grenoble (LAOG) UMR 5571, BP 53,
  38041 Grenoble Cedex 09, France}

\begin{abstract}
  In this article, I examine several observational trends regarding
  protoplanetary disks, debris disks and exoplanets in binary systems
  in an attempt to constrain the physical mechanisms of planet
  formation in such a context. Binaries wider than about 100\,AU are
  indistinguishable from single stars in all aspects. Binaries in the
  5--100\,AU range, on the other hand, are associated with
  shorter-lived but (at least in some cases) equally massive
  disks. Furthermore, they form planetesimals and mature planetary
  systems at a similar rate as wider binaries and single stars, albeit
  with the peculiarity that they predominantly produce high-mass
  planets. I posit that the location of a stellar companion influences
  the relative importance of the core accretion and disk fragmentation
  planet formation processes, with the latter mechanism being
  predominant in binaries tighter than 100\,AU.
\end{abstract}

\keywords{binaries: general --- planetary systems --- planetary systems:
  formation --- planetary systems: protoplanetary disks}

\section{Introduction}

The exponentially growing number of known extrasolar planets now
enables statistical analyses to probe their formation mechanism. Two
theoretical frameworks have been proposed to account for the formation
of gas giant planets: the slow and gradual core accretion model
\citep{lissauer07}, and the fast and abrupt disk fragmentation model
\citep{durisen07}. The debate regarding their relative importance is
still ongoing. Both mechanisms may contribute to planet formation,
depending on the initial conditions in any given protoplanetary disk
\citep[][and references therein]{boley09}. By and large, our
understanding of the planet formation process is focused on the case
of a single star+disk system. Yet, roughly half of all solar-type
field stars, and an even higher proportion of pre-main sequence (PMS)
stars, possess a stellar companion \citep[e.g.,][]{dm91, mathieu94,
  duchene07}. Since the disk and multiplicity phenomena are associated
with similar ranges of distances from the central star, the dynamical
influence of a companion on a disk may be dramatic. Theory and
observations agree that stellar companions can open large gaps in
disks, or truncate them to much smaller radii than they would
otherwise have \citep[e.g.,][]{artymowicz94, ireland08}. The
consequences for planet formation are still uncertain, however.

Observations of protoplanetary disks among PMS stars have revealed
that tight binaries generally show substantially reduced
(sub)millimeter thermal emission \citep{beckwith90, jensen96} as well
as a much rarer presence of small dust grains in regions a few AU from
either component \citep{ghez93, cieza09}. Both trends can be
qualitatively accounted for by companion-induced disk truncation,
which can simultaneously reduce the disk's total mass, outer radius
and viscous timescale. These observational facts have generally been
interpreted as evidence that binaries tighter than $\sim100$\,AU are
much less likely to support gas giant planet formation. However,
follow-up imaging surveys have identified some 50 planet-host stars
that possess at least one stellar companion
\citep[e.g.,][]{patience02, chauvin06, raghavan06, eggenberger07,
  mugrauer09}. In particular, it it is worth noting that about 20\% of
all known planets in binary systems have a stellar companion within
less 100\,AU, so that planet formation in such an environment cannot
be considered a rare occurrence.

In this {\it Letter}, I review several key statistical properties of
PMS and field binary systems that provide insight on the planet
formation process (Sections\,\ref{sec:ci} and \ref{sec:end}). I then
discuss the implications for the main mechanisms of planet formation
in binary systems as a function of their projected separation
(Section\,\ref{sec:implic}). In this study, I only consider binaries
in the 5--1400\,AU separation range, for which current PMS
multiplicity surveys are reasonably complete. The tightest binary
system known to host a planet has a 19\,AU separation. Stellar
companions beyond 1400\,AU are not expected to have much influence on
planet formation.

\section{Planet formation in binaries: initial conditions}
\label{sec:ci}

\subsection{Defining an homogeneous sample of PMS binaries}

In order to draw a broad and homogeneous view of the initial
conditions for planet formation, I compiled a sample of 107 PMS
binaries for which deep (sub)millimeter continuum observations and/or
near- to mid-infrared colors are available in the literature. The
(sub)millimeter data are taken from the work of
\citet{andrews05,andrews07}; for almost all targets, a 1$\sigma$
sensitivity of 15\,mJy or better at 850$\mu$m and/or 1.3mm is
achieved. The median projected separation in this sample is 92\,AU. I
also defined a comparison sample of 222 PMS stars for which no
companion has ever been detected. I focus here on the Taurus and
Ophiuchus star forming regions, the only ones for which
high-resolution multiplicity, photometric and millimeter surveys have
a high completeness rate. The two clouds contribute an almost equal
number of binaries to the sample. Furthermore, both regions have
similar stellar age distributions (median age around 1\,Myr, Ophiuchus
being probably slighter younger on average than Taurus) and their mass
function fully samples the 0.1--1.5\,$M_\odot$ range
\citep[e.g.,][]{luhman99, luhman00}. Finally, Taurus represents an
instance of distributed star formation, while Ophiuchus is a more
clustered environment. These two clouds therefore offer a global view
of the early stages of planet formation among solar-type and
lower-mass stars.

\subsection{Inner disk presence and survival}

I first address the question of the presence of dust in the
planet-forming region, namely the innermost few AU around each
component, within binary systems. To probe the presence of an
optically thick dusty inner disk, I used near- to mid-infrared
colors. I selected the following standard thresholds to conclude that
a circumstellar disk is present: $[3.6]-[8.0] \ge 0.8$\,mag, $K-N \ge
1.75$\,mag, $K-L \ge 0.35$\,mag, $\alpha_{2-14\mu\mathrm{m}}>-1.7$
\citep[e.g.,][]{cieza09, mccabe06, bontemps01}. About 80\% of the PMS
binaries considered here have {\it Spitzer}/IRAC colors, which are
used whenever available.

\begin{figure}
\plotone{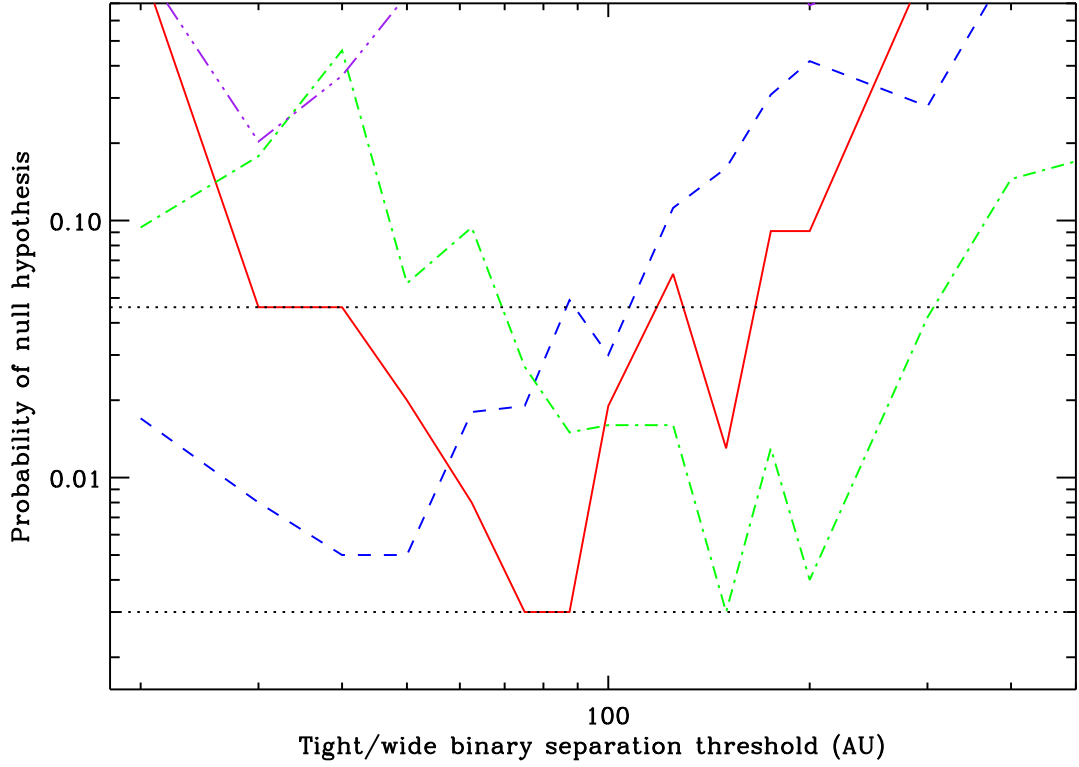}
\caption{Probability of several null hypotheses that compare the
  observed properties of tight and wide binaries as a function of the
  radius used to separate both categories. The following null
  hypotheses are tested: planets in tight and wide binaries have the
  same frequency of high mass planets, namely above the median for all
  known extrasolar planets ({\it red solid curve}); tight and wide
  binaries have the same probability of hosting optically thick dust
  within a few AUs of the central stars ({\it blue dashed curve}); the
  distributions of 850$\mu$m fluxes for tight and wide binaries with
  evidence of optically thick inner disks are indistinguishable ({\it
    green dot-dashed curve}); the frequency of the debris disk
  phenomenon is the same for tight and wide binaries ({\it purple
    triple-dot-dashed curve}). The horizontal dotted lines represent
  the 2 and 3$\sigma$ confidence levels (top and bottom,
  respectively). Notwithstanding small number statistics (responsible
  for the jagged appearance of the curves), the first three null
  hypotheses can be rejected at the 99.5\% confidence level. Only the
  debris disk phenomenon is independent of the binary
  separation. \label{fig:proba}}
\end{figure}

\cite{cieza09} have demonstrated that tighter binaries have a much
lower probability of hosting circumstellar dust. The same effect is
observed here in a somewhat smaller sample. The median separation of
binaries with an inner disk in this sample is about 100\,AU, whereas
that of disk-free binaries is 40\,AU. The simplest interpretation of
this trend is that disks in tight binaries are dissipated much faster
than in wide systems \citep[][Kraus et al., in prep.]{cieza09}. To
extend upon this previous analysis, I used the two-sided Fischer exact
test to determine the probability that wide and tight binaries have a
different proportion of diskless systems, using a sliding threshold to
split the sample. As shown in Figure\,\ref{fig:proba}, the difference
is significant at the 2$\sigma$ level or higher for a wide range of
threshold separations. In particular, this analysis reveals that {\it
  the observed reduced disk lifetime in binaries only applies to
  systems that are tighter than about 100\,AU}. On the other hand,
there is no statistical difference between binaries wider than 100\,AU
and single stars.

\subsection{Total mass of the initial dust reservoir}

While near- and mid-infrared emission best traces the presence of dust
within a few AU of star, only long-wavelength flux measurements can
probe the total dust mass of protoplanetary disks
\citep[e.g.,][]{beckwith90}. From the sample defined above, I selected
those objects which show evidence of an optically thick inner disk (as
defined above) and have been observed in the (sub)millimeter. The
median separation in this subsample of 44 binaries is 130\,AU. While
the 850$\mu$m survey of Ophiuchus is not yet as complete as that of
Taurus, the existing 1.3mm observations of PMS stars are generally
less sensitive to cold dust. Since using both wavelengths yield
similar conclusions but with lower significance for the 1.3mm one, I
focus here on 850$\mu$m measurements.

As has long been known, tight binaries have a different distribution
of submillimeter fluxes than wide ones, with a much lower median flux
(13\,mJy vs 50\,mJy at 850$\mu$m using a 100\,AU separation threshold)
and only very few high-flux systems \citep{jensen96, andrews05}. I
compared the distributions of 850$\mu$m fluxes for tight and wide
binaries defined by the same sliding threshold as above using the
conservative survival analysis Peto-Pentrice Generalized Wilcoxon test
to account for upper limits. I find that wide and tight binaries are
different at the 2$\sigma$ level or higher if the separation threshold
is in the 75-300\,AU range (see Figure\,\ref{fig:proba}).
I therefore conclude that {\it binaries with a projected separation
  smaller than 300\,AU have a substantially reduced submillimeter
  flux}. On the other hand, the distribution of 850$\mu$m fluxes for
wide binaries is indistinguishable from that of single stars.

\begin{figure}
\plotone{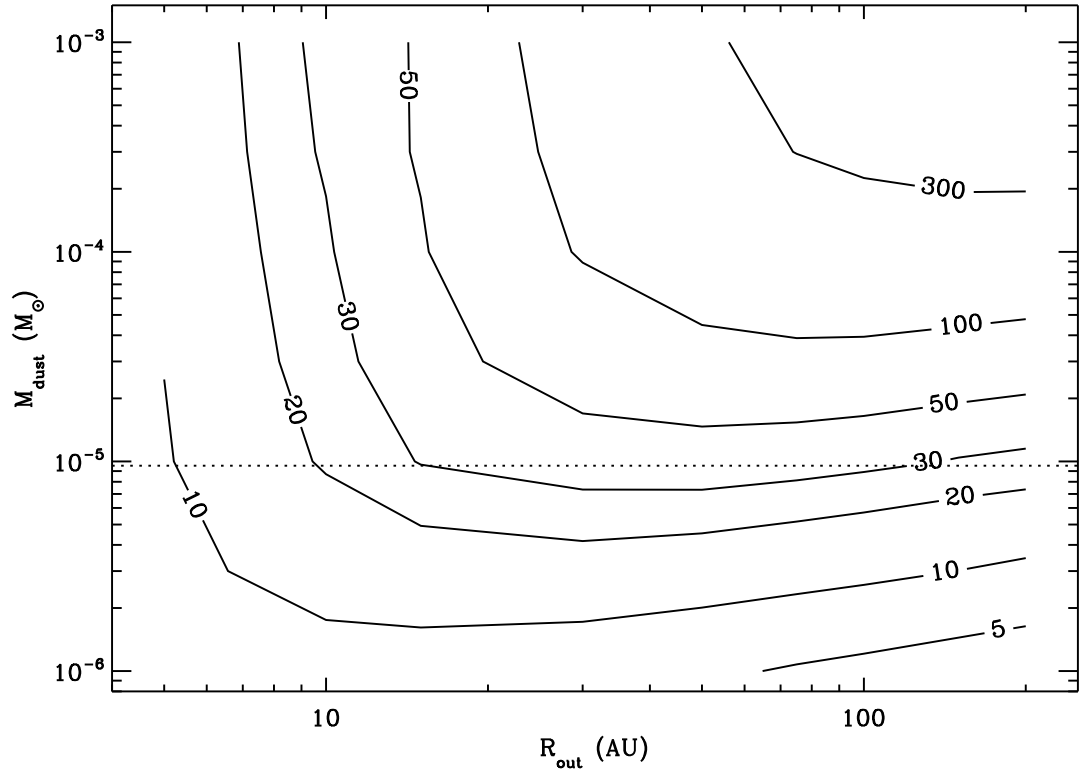}
\caption{
  Model 850$\mu$m flux (in mJy) for a typical flared disk (see model
  parameters in section 2.3) with varying total dust mass $M_{dust}$
  and outer radius $R_{out}$, viewed at an inclination of 60\degr (see
  Section 2.3 for more detail). The horizontal dotted line represent a
  total disk mass $M_{d}=M_J$, assuming a 100:1 gas-to-dust
  ratio. Compact disks, with $R_{out} \lesssim 10$\,AU must be massive
  to yield $F_{850\mu\mathrm{m}} \gtrsim
  20$\,mJy.\label{fig:diskmass}}
\end{figure}

In past studies, it has been assumed that a reduced (sub)millimeter
flux necessarily implies a reduced total dust mass independently of
the disk properties \citep[for instance, see the prescription used
by][]{andrews05}. While this is true in general, it is unclear whether
this assumption is valid for severely truncated disks for which
optical depth effects may become important. The model constructed by
\cite{jensen96} seems to support this hypothesis, but these authors
assumed that tight binaries are always surrounded by a massive
circumbinary structure, which we now know is rare. To revisit this
issue, I have computed a grid of radiative transfer models using the
MCFOST code \citep{pinte06} to compute the 850$\mu$m flux of a disk
with a typical $\Sigma(r) \propto r^{-1}$ surface density profile, an
0.1\,AU inner radius and a flaring power law $H(r)\propto
r^{1.125}$. Emission from the central star is modeled as a 4000\,K,
2$L_\odot$ photosphere and a distance of 140\,pc is assumed. The dust
is assumed to be made of astronomical silicates with a $a^{-3.5}$
power law size distribution ranging from 0.03$\mu$m to 1\,mm. The only
variables in the model are the disk outer radius, $R_{out}$, and the
total dust mass, $M_{dust}$. Figure\,\ref{fig:diskmass} demonstrates
that the proportionality between total dust mass and submillimeter
flux observed for large disks breaks down for $R_{out} \lesssim
30$\,AU as the disk becomes optically thick to its own emission.

Disk truncation by an outer stellar component is dependent on the
orbital parameters and mass ratio of the binary system
\citep{artymowicz94}. It is therefore not possible to uniquely
associate a binary separation with a tidally-set value of
$R_{out}$. The ratio between these quantities is typically in the
broad 2.5--5 range. Systems whose separation is less than 100\,AU are
therefore expected to possess disks whose outer radius is 40\,AU or
less. In this configuration, total disk masses of at least $M_{J}$ are
necessary to produce 850$\mu$m fluxes as low as $\sim20$--30\,mJy. In
the sample studied here, about a third (6 out of 19) of all binaries
that are tighter than 100\,AU and possess an inner disk have an
850$\mu$m flux that is higher than 30\,mJy. Therefore, {\it a
  significant fraction of the circumstellar disks in tight binaries
  ($\lesssim 100$\,AU) are massive enough to potentially form gas
  giant planets, despite their much lower (sub)millimeter fluxes.}

\section{Planet formation in binaries: end results}
\label{sec:end}

\subsection{Distribution of planetary masses}

Les us turn our attention to mature planetary systems. As of this
writing, there are 38 exoplanets that are in a system with at least 2
stellar components (using 1400\,AU as the upper limit for binary
separation), including 5 systems with a stellar companions within
25\,AU \citep[][and references therein]{mugrauer09}. Most of these
planet-bearing stars are of solar type. The overall detection rates of
gas giant planets in binary systems and in single stars are
undistinguishable \citep{bonavita07}. There is marginal evidence that
planets in binaries tighter than $\sim100$\,AU may be somewhat less
frequent than one would assume based on the frequency of planets in
wider binaries \citep[by 6.0$\pm$2.7\%,][]{eggenberger08}. However,
the small sample size, adverse selection biases and incompleteness of
current multiplicity surveys are such that it is premature to reach
definitive conclusions. In any case, we can use this sample to test
whether the separation of the stellar binary has any influence on
planet properties.

\begin{figure}
\plotone{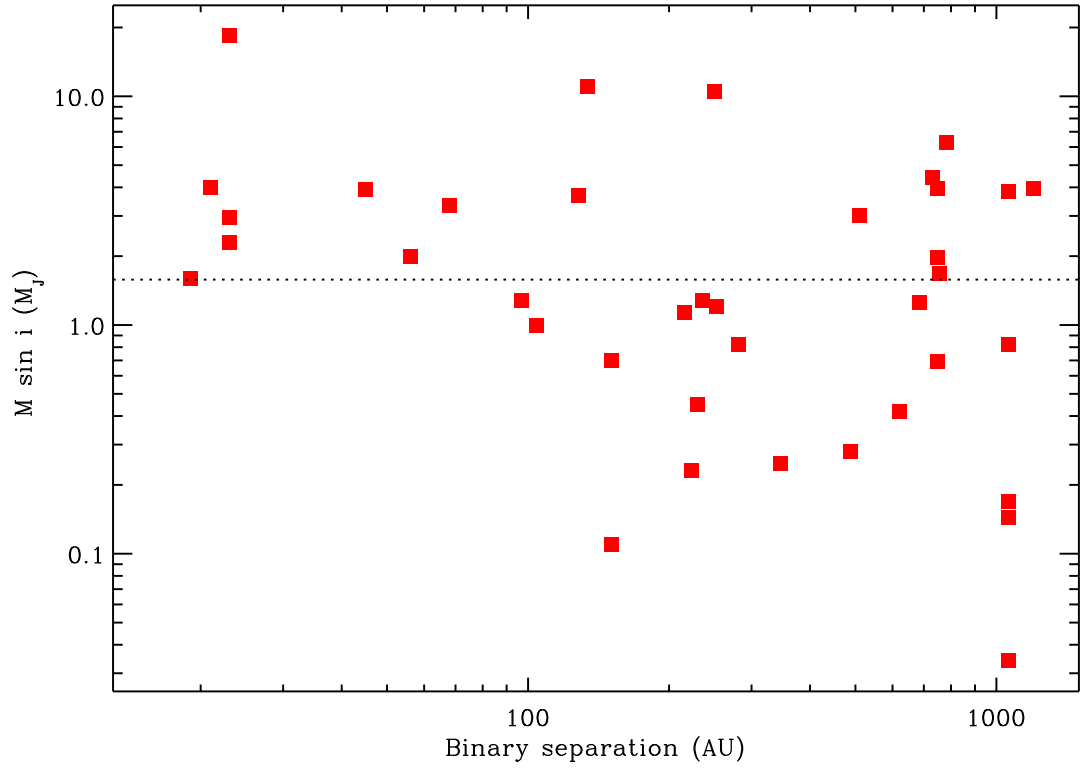}
\caption{Distribution of masses ($M \sin i$) for planets in binary
  systems as a function of the binary separation. Note the deficit of
  planets with $M\sin i \lesssim M_J$ for binaries tighter than
  100\,AU. The horizontal dotted line represent the median mass for
  all known extrasolar planets to date (including among single
  stars).\label{fig:planets}}
\end{figure}

Despite an earlier claim for a distinct period-mass distribution
\citep{zucker02}, \cite{bonavita07} have shown that there is
essentially no difference in the properties of planetary systems
around single stars and in binary systems. However, a previously
unrecognized trend is evident in Figure\,\ref{fig:planets}. While
planets covering two orders of magnitude in mass can be found in wide
binaries (as around single stars), systems tighter than $\sim100$\,AU
appear to host only high-mass, $M \gtrsim M_{J}$, planets. To quantify
this effect, I used the two-sided Fischer Exact test to determine
whether close and wide binaries (with the usual sliding threshold)
have different proportion of high- and low-mass planets. I used
1.6$M_J$, the median for all planets known to date\footnote{This is
  also the median for the subsample of all planets within multiple
  stellar systems.}, to separate low- from high-mass
planets. Figure\,\ref{fig:proba} confirms that {\it binary systems
  tighter than about 100\,AU produce a distribution of planets that is
  strongly biased towards the highest masses.} This conclusion is
significant at the 3$\sigma$ level. 

It is important to test whether this trend is not a mere consequence
of a selection bias, as a close stellar companion can alter the
detectability of a planet-induced radial velocity signal. To evaluate
this possibility, I build on the ``uniform detectability'' sample
defined by \cite{fischer05} which contains all stars for which
close-in planets as low mass as 0.3$M_J$ (well below the apparent
cut-off in mass for planets in tight binaries), as well as 1$M_J$
planets on a 4yr orbit, could be detected. In the current sample of
binary planet hosts, the proportions of stars that belong to the
uniform detectability sample among binaries tighter and wider than
100\,AU are indistinguishable (5/9 and 24/40, respectively). I
therefore conclude that the trend discussed above is unlikely to be
the consequence of a selection bias or of observational limitations.

\subsection{Occurrence of planetesimal disks}

An indirect signpost of planet formation is the debris disk
phenomenon. In these systems, small dust grains are produced via the
collisions of large solid bodies
\citep{zuckerman01}. \cite{trilling07} observed 69 A- and F-type known
binaries with {\it Spitzer} and found debris disks in systems spanning
6 decades in separation. They further suggested that intermediate
separation (3--30\,AU) binaries are substantially less likely to host
a debris disks than either tighter or wider systems, although the
formal significance of this difference is marginal at best. No such
trend was found by \cite{plavchan09}, who included 24 A- through
M-type binaries in their own {\it Spitzer} survey. This latter survey
focused on targets that are more similar in mass to exoplanet hosts
and the PMS population discussed in the previous section.

I used the two-sided Fischer test to determine whether the occurrence
of debris disks is indeed different in tight and wide binaries, using
the same sliding threshold as above (see
Figure\,\ref{fig:proba}). There is no significant difference for any
value of the threshold in the sample from \cite{trilling07}, nor in a
combined sample that also includes systems from \cite{plavchan09}. The
combined sample contains 52 binaries in the 5--1400\,AU range, with a
median separation of 50\,AU, an increase of 15 sources from the sole
sample of \citeauthor{trilling07}. In addition, the occurrence rates
of debris disks in binary systems and single stars are very similar
\citep{trilling07}. In other words, {\it any 0.5--2$M_\odot$ star,
  irrespective of the presence of a companion (within the 5--1400\,AU
  range studied here), may experience the early phases of planet
  formation up to the planetesimal stage}.

\section{Discussion and implications}
\label{sec:implic}

This analysis has revealed a clear dichotomy between tight and wide
binaries. Systems with separation $\gtrsim 100$\,AU are
indistinguishable from single stars as far as the initial conditions
and end product of planet formation are concerned. The only caveat to
this statement is the possibility of mild disk truncation in
100-300\,AU systems, but most disks in these systems retain a mass
reservoir that is sufficient to build up gas giant planets. On the
other hand, planet formation in binaries with separations $\lesssim
100$\,AU is characterized by a much shorter clearing timescale for the
protoplanetary disks and a strong bias towards high-mass
planets. Despite these differences, planetesimals and mature planetary
systems appear to form at roughly the same frequency as around other
stars. Furthermore, while protoplanetary disks are more compact in
tight systems because of truncation, a significant fraction of them
possess large mass reservoirs (at least several times $M_J$). Taken
together, {\it these results suggest that planet formation in binaries
  tighter than 100\,AU proceeds through a different, but not much less
  frequent, mechanism compared to wide binaries and single stars.}

The shorter disk lifetime in tight binaries makes it extremely
difficult to form gas giant planets through the core accretion model,
especially if the final planets are particularly massive. Rather, this
combination of observed trends supports an abrupt process to form
planets in tight binaries, such as the disk fragmentation
model. Indeed, this mechanism can be extremely efficient in the case
of a compact, massive protoplanetary disk which is naturally prone to
gravitational instability. Furthermore, gravitational perturbations
induced by a close stellar companion can trigger the instability even
though the disk itself is not unstable to its own gravity
\citep{boss06}. On the other hand, considering the long survival
timescale and slim chances of gravitational instabilities, disks
located within wide binaries and around single stars are good
candidates to form planets via the core accretion model in their inner
regions \citep[e.g.,][]{boley09}.

While a violent process is most likely responsible for the formation
of planets in tight binaries, it is however unclear whether all
planets in wide binaries form through a single mechanism. Indeed, it
is also conceivable that high-mass planets ($\gtrsim M_J$) mostly form
via disk fragmentation, while lower mass planets are preferentially
the result of core accretion. This scenario would naturally alleviate
the difficulty of the core accretion model to form the highest-mass
planets in less than a few Myr. This hypothesis has the additional
advantage that it could also apply to tight binaries. Indeed, since a
stellar companion located within less than 100\,AU dramatically
shortens the disk lifetime, core accretion is essentially prevented
from occurring, accounting for the absence of low-mass ($< 2 M_J$) gas
giant planets in tight binaries. Planetesimals can presumably form in
either scenario, accounting for the observations regarding the debris
disks phenomenon. In summary, it remains to be determined whether the
trends discussed here indicate an actual dichotomy between the main
planet formation theories or a mere change of the relative importance
of the two models as a function of the location of the stellar
companion. Improving the statistical significance of the various
trends discussed here and determining the exact properties of disks
within tight PMS binaries will help shed further light these two
possibilities.


\acknowledgments I am grateful to Silvia Alencar and Jane
Gregorio-Hetem for organizing and inviting me to ``Special Session 7''
at the IAU 27th General Assembly held in Rio de Janeiro, where this
work was first presented, as well as to Anne Eggenberger, Deepak
Raghavan, David Rodriguez and Peter Plavchan for invaluable input
regarding exoplanets and debris disks. The work presented here has
been funded in part by the Agence Nationale de la Recherche through
contract ANR-07-BLAN-0221.

\clearpage

\begin{table*} 
  \caption{\label{tab:proto}Sample of Taurus and Ophiuchus protoplanetary disks considered in this study.}
\begin{tabular}{llcccccl}
\hline
Target & Alt. Name &  Sep. & IR color\tablenotemark{a} & Inner disk? &
$F_{850\mu\mathrm{m}}$ & $F_{1.3\mathrm{mm}}$ & References\tablenotemark{b} \\
 & & [AU] & [mag] & & [mJy] & [mJy] & \\
\hline
\multicolumn{8}{c}{Taurus-Auriga}\\
\hline
CZ Tau & & 46 & 3.41 & Y & $<9$ & $<30$ & 1,2 \\
DD Tau & & 79 & 1.99 & Y & $<42$ & 17 & 1,2 \\
DF Tau & & 13 & 1.37 & Y & 8.8 & $<25$ & 3,2 \\
DI Tau & & 17 & 0.76 & N & -- & -- & 3 \\
DK Tau & & 350 & 1.27 & Y & 80 & 35 & 3,2 \\
\hline
\end{tabular} 
\tablenotetext{a}{Unless otherwise indicated by a note, the infrared
  color is the {\it Spitzer}/IRAC $[3.6]-[8.0]$ color.}
\tablenotetext{b}{References: 1) \cite{luhman06}; 2) \cite{andrews05};
  3) \cite{hartmann05}; 4) \cite{cieza09}; 5) \cite{mccabe06}; 6)
  \cite{guilloteau99}; 7) \cite{kenyon95}; 8) \cite{duchene10}; 9)
  \cite{jensen03}; 10) \cite{jensen96}; 11) \cite{andrews07}; 12)
  \cite{bontemps01}.}
\end{table*}

\clearpage


\end{document}